# Reconstructing Global Daily CO$_2$ Emissions via Machine Learning

Tao Li[1], Lixing Wang[1], Zihan Qiu[2], Philippe Ciais[3], Taochun Sun[1], Matthew W. Jones[4], Robbie M. Andrew[5], Glen P. Peters[5], Piyu ke[1], Xiaoting Huang[1] Robert B. Jackson[6] and Zhu Liu[1*]

[1]Department of Earth System Science, Tsinghua University, Beijing, 100084, China

[2]Tsinghua University Institute for Interdisciplinary Information Sciences (IIIS) FIT Building, Tsinghua University, Beijing 100084, China

[3]Laboratoire des Sciences du Climat et de l'Environnement, IPSL, CEA/CNRS/UVSQ, Université Paris-Saclay, Pairs, 91191, France

[4]Tyndall Centre for Climate Change Research, School of Environmental Sciences, University of East Anglia, Norwich Research Park, Norwich NR4 7TJ, UK

[5]CICERO Center for International Climate Research, Oslo 0349, Norway

[6]Earth System Science Department, Stanford University, 473 Via Ortega, Stanford, CA 94305, USA.

*Email: zhuliu@tsinghua.edu.cn

## Abstract

High temporal resolution CO$_2$ emission data are valuable for understanding the drivers of emission changes. Current emission datasets, however, are generally only available with an annual resolution. Here, we extended a global daily CO$_2$ emissions dataset backwards in time to 1970 using a machine learning algorithm, which was trained to predict historical daily emissions on national scales based on relationships between daily emission variations and predictors established for the period since 2019. Variation in daily CO$_2$ emissions far exceeded the smoothed seasonal variations. For example, the range of daily CO$_2$ emissions of China and India increased from 1.2 and 0.2 Mt/day in 1970 to 10.8 and 4.2 Mt/day in 2022, reaching approximately 31% of the year average daily emissions of China and 46% of India in 2022, respectively. The relationship between daily CO$_2$ emission and the ambient temperature is well described by a linear-plus-plateau function, in which we identified the emission-climate tipping temperature ($T_c$) is 16.9°C for global average (19.5°C for China, 15.0°C for U.S., and 18.2°C for Japan), demonstrating increased emissions associated with higher ambient temperature. The long-term time series spanning over fifty years of global daily CO$_2$ emissions reveals an increasing trend in emissions due to extreme temperature events, driven by the rising frequency of these occurrences. This work adds to evidence that, due to climate change, greater efforts may be needed to reduce CO$_2$ emissions.

# Main

Daily emission data are required to study the impact of short-term temperature fluctuations, such as heat waves lasting several days to weeks, on $CO_2$ emissions. High-temporal resolution estimates of $CO_2$ emissions provide a more detailed picture and offer insights into driving factors that annual data may overlook. Utilizing high-resolution $CO_2$ emission estimates, the effects of extreme temperature events, COVID-19, as well as blackouts on $CO_2$ emissions can be quantified, enhancing the understanding of emission change drivers[1]. High temporal resolution $CO_2$ emissions inventory can be used in atmospheric inversion models to quantify the impact of extreme temperature events on forest carbon sinks[2]. These assessments are vital for formulating policies aimed at achieving net-zero greenhouse gas emissions through forest carbon sinks.

Fossil CO2 emissions arise predominantly from the combustion of fossil fuels linked to activities such as electricity generation, mobility, industrial production, and similar. With various sources of activity data and satellite measurements, some data can be obtained at daily or even hourly or sub-hourly frequency, suggesting the possibility of presenting $CO_2$ emissions with high temporal resolution. For example, attempts have been made to estimate the $CO_2$ emissions decline due to COVID-19 based on forced confinement policies[3,4] and data on mobility changes[5]. Extending this further, if the high resolution data is also collected and processed in near real time [6,7] then the amount and dynamics of activity in certain sectors can be used to construct the $CO_2$ emissions estimates in near-real-time and at high temporal resolutions[8]. Satellite $NO_2$ measurements can be used to build top-down methods for estimating low latency $CO_2$ emissions for country[9] and constrain the $CO_2$ emission reduction from a bottom-up datase[10]. However, current high temporal resolution $CO_2$ emission datasets are only available for recent periods and remain inaccessible for historical periods preceding this time frame.

Using scale factors from the Temporal Improvements for Modeling Emissions by Scaling (TIMES)[11] or the Emissions Database for Global Atmospheric Research (EDGAR) temporal profiles library (hereafter "EDGAR_profile")[12], annual or monthly $CO_2$ emissions can be distributed daily. International agencies and organizations such as International Energy Association (IEA), Carbon Dioxide Information Analysis Center (CDIAC), Open-source Data Inventory for Anthropogenic $CO_2$ (ODIAC), EDGAR, Global Carbon Budget (GCB), and more officially National Inventory Reports can provide the annual or monthly fossil fuel $CO_2$ emission data globally[13-17]. TIMES assigns fixed scale factors for different days of the week by country. For example, in the U.S., the scaling factors for Sunday, Saturday, and weekdays are 0.913268, 0.962927, and 1.024760, respectively. EDGAR_profile provides scale factors for each day of week by sector. These factors primarily reflect variations in daily $CO_2$ emissions associated with the day of week, yet due to the downscaling measure from monthly or yearly dataset, such approaches are still unable to reflect the fine temporal resolution environmental influences such as temperature changes. The methods also often use generic scaling factors across countries and years. For example, fluctuations in temperatures within a few days can alter heating requirements in winter[18] and cooling demands in summer[1], both of which significantly affect $CO_2$ emission levels but are difficult to be captured by TIMES and EDGAR_profile.

Here we estimated historical global daily $CO_2$ emissions in 1970-2018 by training eXtreme Gradient Boosting (XGBoost)[19] models to predict daily $CO_2$ emissions in the period 2019-2022, when high-resolution data already exist[6,8]. The machine learning approach includes the impact of daily variations in ambient temperature and other factors, by considering the non-linear effect of temperature on $CO_2$ emissions and their interactions with temporal surrogate variables (e.g., day of week and month of year). Quantile regression was applied to assess data uncertainty, and the model performance was evaluated using the 10-fold cross validation method and back-extrapolation validation. The back-extrapolation validation was implemented based on $CO_2$ emissions in the United States as estimated by Vulcan[20] from 2010 to 2015. The reconstructed global daily $CO_2$ emission dataset is useful for assessing the impacts of extreme temperature on $CO_2$ emissions and informing-policy making. It also provides data support to chemical transport models, which is critical for evaluating the effects of extreme temperature on forest carbon sinks.

## Results
**Daily $CO_2$ emissions from 1970 to 2022**
The global daily $CO_2$ emissions increased consistently from 1970 to 2022, with the annual averages increasing from 50.6 Mt/day in 1970 to 106.9 Mt/day in 2022 (Figs. 1 and S1). Among them, China and India experienced substantial increases in $CO_2$ emissions, while $CO_2$ emissions in the U.S., Germany, and other countries remained relatively stable. In 1970, the daily average $CO_2$ emissions in China and India were 2.9 Mt/day and 0.6 Mt/day, respectively. By 2022, daily average $CO_2$ emissions had increased to 35.0 Mt/day for China and 9.2 Mt/day for India, representing increases of 12.1 times and 15.3 times, respectively. The daily $CO_2$ emissions of the U.S., Japan, and Italy peaked around 2007 after experiencing growth since 1970. Germany's $CO_2$ emissions peaked much earlier, around 1990, corresponding to reunification and subsequent economic changes.

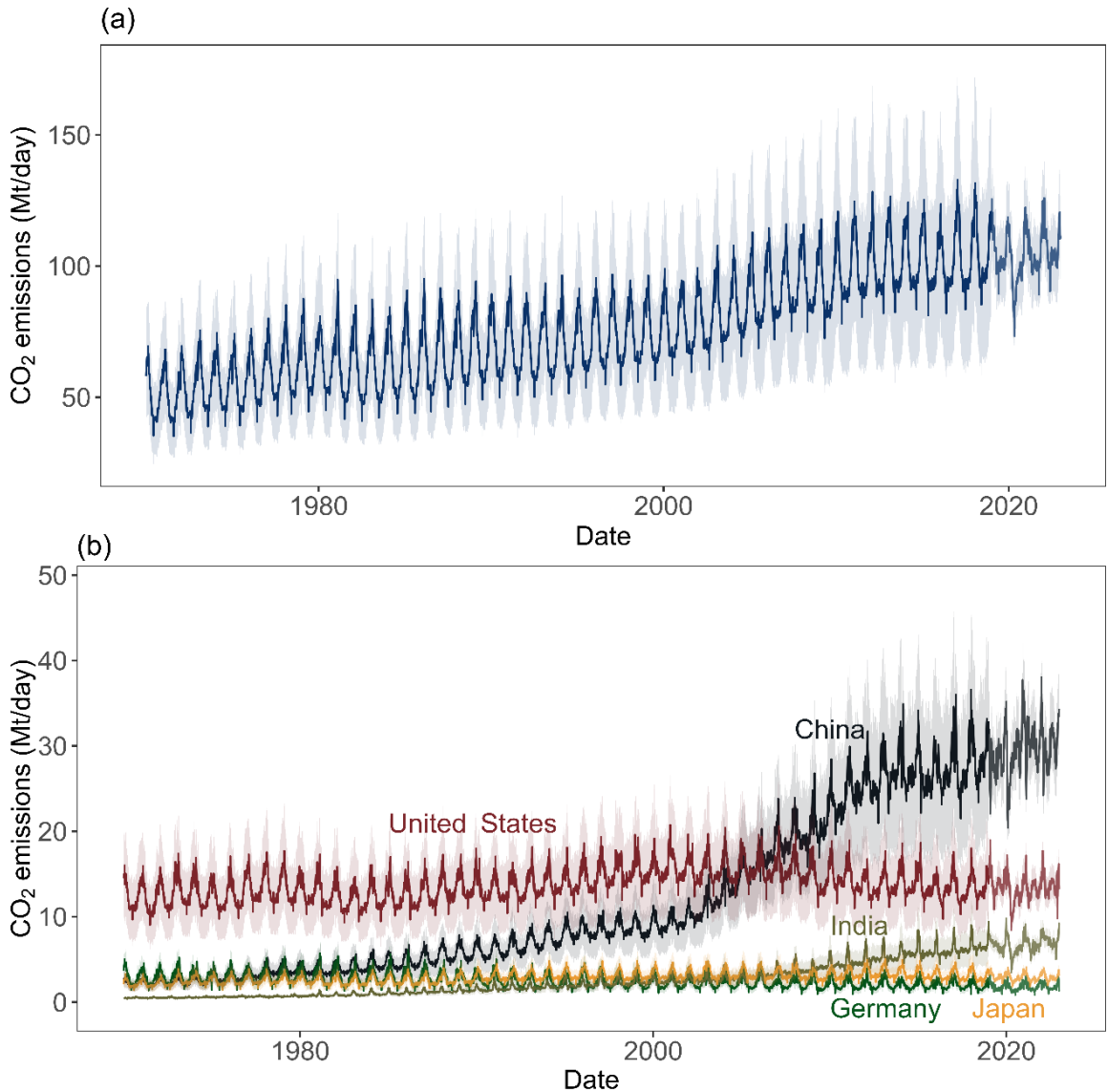

**Fig. 1. Daily CO$_2$ emissions from 1970 to 2022 for (a) global and (b) individual countries.** The shaded areas indicate the uncertainty of CO$_2$ emissions.

The reconstructed daily CO$_2$ emissions of various countries from 1970 to 2022 showed significantly higher variability compared to the annual average CO$_2$ emissions estimated by Global Carbon Budget 2022[18] (Figs. 1, S1 and S2). The global daily emissions peaked around December and were lowest around June in each year, reflecting the increasing energy consumption due to heating or cooling demand and associated changes in human activities. The peak-to-valley difference of CO$_2$ emissions within a year (maximum CO$_2$ emissions minus minimum CO$_2$ emissions) was around 32 Mt/day in 1970 (equivalent to X% of daily average CO2 emission) and increased to 35 Mt/day in 2022 (30% of daily average CO$_2$ emission). China and India especially showed significant increases in their peak-to-valley differences from 1970 to 2022, rising from 1.2 Mt/day to 10.8 Mt/day for China and from 0.2 Mt/day to 4.2 Mt/day for India. The peak-to-valley differences in countries like the U.S. and Japan changed relatively little, from 5.8 and 1.8 Mt/day in 1970 to 7.9 and 2.0 Mt/day in 2022. Given the significant variability in daily CO$_2$ emissions, India's peak emission in 2022 (10.3 Mt/day) nearly surpassed the valley emission of the U.S. (9.4 Mt/day), even though India's annual average CO$_2$ emissions (7.3 Mt/day) in 2022 were approximately half of the U.S.'s annual average $CO_2$ emissions (13.6 Mt/day). China's annual average $CO_2$ emissions surpassed those of the U.S. around 2010, but its peak daily $CO_2$ emissions had already surpassed the U.S.'s valley daily $CO_2$ emissions in 2005.

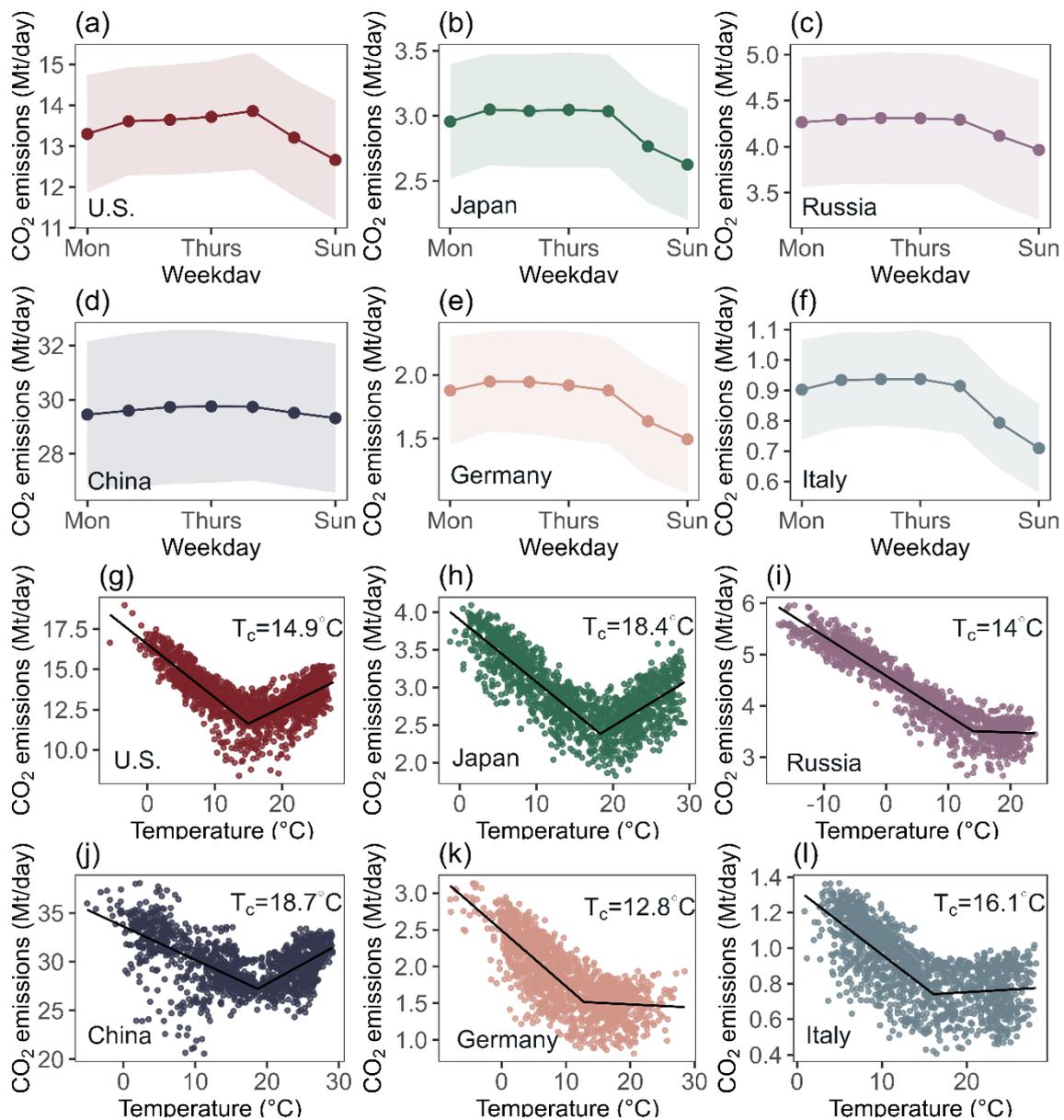

**Fig. 2. Weekly average of $CO_2$ emissions and the relationship between daily $CO_2$ emissions and population-weighted temperature for major countries.** (**a-f**) are the weekly average of $CO_2$ emissions during 2019 and 2022, while (**g-l**) display the relationship between daily $CO_2$ emissions and population-weighted temperature. Note that in (g-l), data records for public holidays have been removed. $T_c$ represents critical emission-climate temperature.

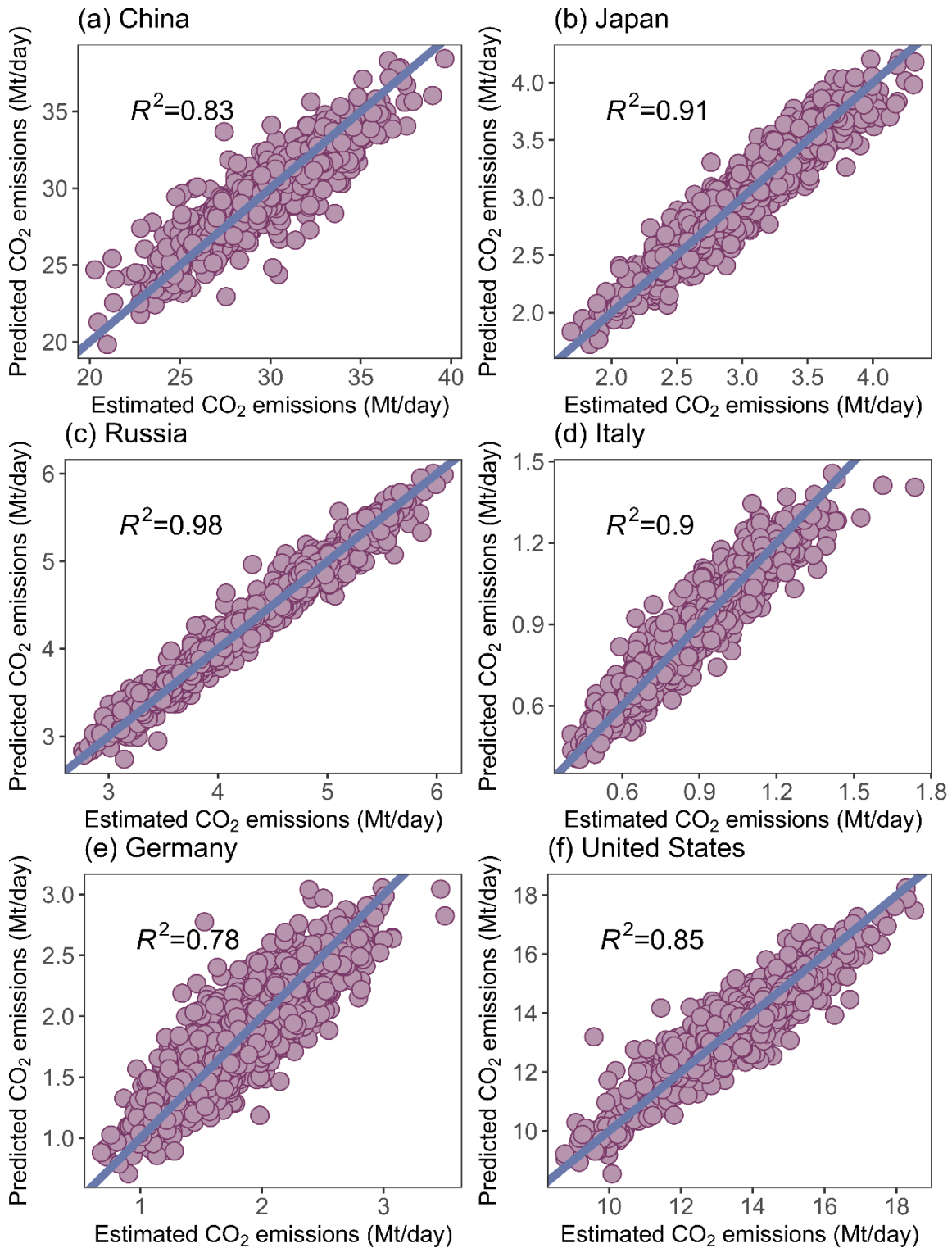

**Fig. 3. Predictive performance of the XGBoost models used in reconstructing the daily $CO_2$ emissions for (a) China, (b) Japan, (c) Russia, (d) Italy, (e) Germany, and (f) the United States.** The blue line is the 1:1 line and the red solid line is the linear fitting line. The Slope is for the regression line. Refer to Methods for the calculations of $R^2$.

**Drivers of daily $CO_2$ emission variations**

We used interpretable machine learning (See Methods) to identify the drivers of daily $CO_2$ emission variations. These predictor variables were also used for reconstructing daily $CO_2$ emissions including population-weighted temperature, day of week, day of month, month of the year, and whether it is a public holiday. The impact of daily temperature on emissions is related to the energy consumption associated with heating and cooling demands, typically with a critical temperature range for heating between 10-20°C (Fig. 2). We found that the relationship between daily $CO_2$ emission and the ambient temperature is well described by a linear-plus-plateau function, as shown in Fig. 2. A negative correlation observed between daily $CO_2$ emission and ambient temperature before this critical temperature and a positive correlation after it. The critical temperatures for the U.S., China, and Japan are 14.9, 18.7, and 18.4°C, respectively. Even excluding the residential sector where $CO_2$ emissions were calculated according to heating and cooling degree days, the critical temperatures were also displayed (Fig. S3). The critical temperature reflects the increased energy consumption due to heating and cooling demand, and varies across countries depending on climate, infrastructure development, income level and other factors. The relationship derived for China is weaker, potentially due to geographic divergence in this relationship; analysis at provincial level would likely give stronger results and will be investigated in future work. Further, these relationships will change not just geographically but also over time, for example as increased wealth leads to increased ability to pay for air conditioning.

Due to reduced operating hours of factories and retail outlets and reduced transportation demand on weekends, $CO_2$ emissions levels on Saturdays and Sundays were significantly lower than on weekdays in most major countries. Italy showed the most pronounced reduction, with Sunday $CO_2$ emissions 20.7% lower than the average level (Monday to Sunday) and Saturday emissions 11.4% lower. The reduction in China was less noticeable, with Sunday $CO_2$ emissions 0.9% lower than the average, possibly due to its intense work environments. Additionally, the day of the month and month were chosen to indicate week-to-week differences and seasonal variations.

These selected variables effectively indicated daily emission variations. Based on 10-fold cross-validation, the model's predictive performance $R^2$ ranged from 0.78 (Germany) to 0.97 (Russia), with an average of 0.86 (Fig. 3). We then calculated the Shapley index values (SHAP) to assess the variable importance, i.e., the contribution of the predictor variables in driving daily $CO_2$ emissions. SHAP is a model-agnostic method commonly used to evaluate variable contributions[22-24]. For each predictor variable, SHAP values are calculated by assessing the change in model predictions when the variable is included and excluded, across all possible combinations of the other predictor variables[24]. The mean absolute SHAP value for a predictor variable represents its overall importance in influencing model predictions; higher values indicate a greater impact on predictions. To enhance interpretability, we scaled the mean absolute SHAP values so that the total importance of all predictor variables equals 100%.

According to scaled mean absolute SHAP values, temperature and the day of the week were identified as the most important variables in driving daily $CO_2$ emission variations across countries (Fig. S4). Except for Italy, temperature was the most important variable for all countries, with scaled mean absolute SHAP values ranging from 34.2% (Japan) to 40.1% (U.S.), highlighting its significant impact on daily $CO_2$ emissions. In Italy, the day of the week was the most important variable, with a scaled mean absolute SHAP value of 42.2%, while temperature was the second most important variable, with a scaled mean absolute SHAP value of 28.3%. Italy's daily $CO_2$ emissions showed a relatively higher reduction on Sunday, which was 20.7% lower than the average $CO_2$ emissions of other days. In contrast, China displayed lower reductions on Sundays, being only 0.9% lower than the averages of other days. Therefore, the day of the week (in China?) showed much lower scaled mean absolute SHAP values (10.9%), making it the least important variable (excluding the variable of is a holiday).

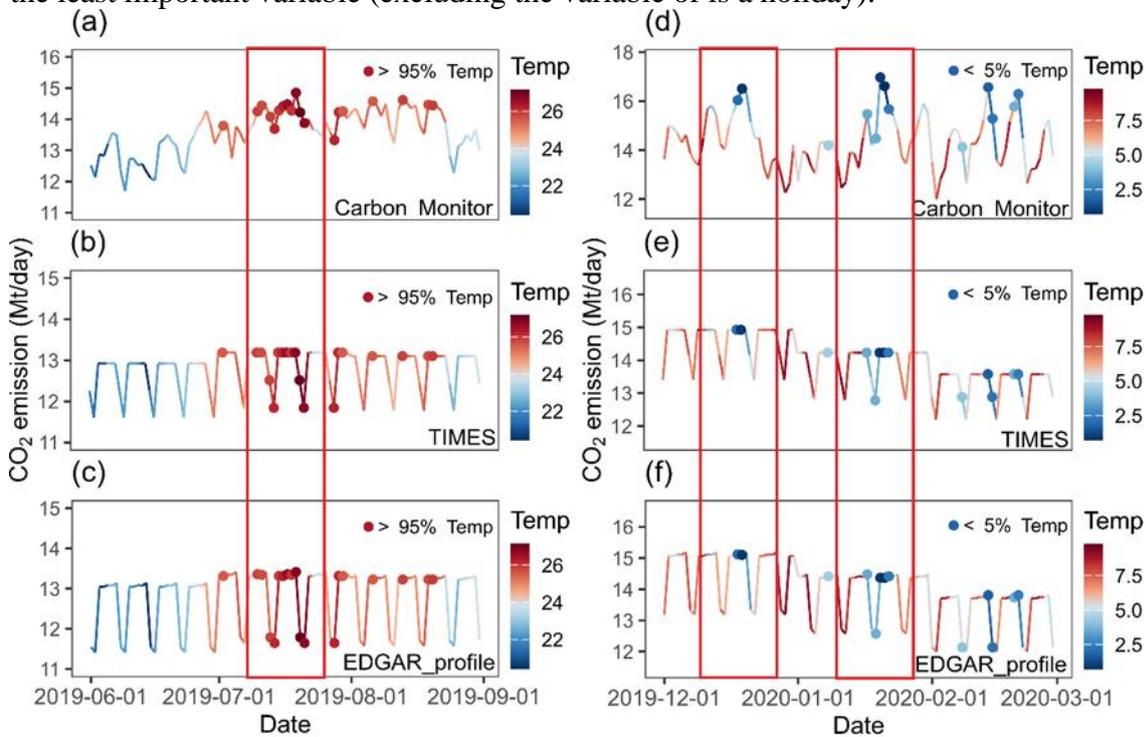

**Fig. 4. Daily CO₂ emissions and population-weighted temperature (Temp) in the United States during 2019 and 2020. a**, **b**, and **c** indicate the CO₂ emissions in summer (June, July, and August), while **d**, **e**, and **f** indicate the CO₂ emissions in winter (December, January, and February). **a** and **d** are the emissions estimated by Carbon Monitor; **b** and **e** are distributed by TIMES; **c** and **f** are distributed by EDGAR_profile. The temporal variation lines are colored according to population-weighted temperature. Blue and red dots represent temperatures falling below the 5th percentile and above the 95th percentile, respectively, of the population-weighted temperatures over 53 years (1970-2022).

**Extreme temperature increases CO₂ emissions**

We found that CO₂ emissions during extreme cold (defined by single days with population-weighted temperature below the 5th percentile over 1970-2022) and hot days (defined by single days with population-weighted temperature above the 95th percentile over 1970-2022) were significantly higher than CO₂ emissions during non-extreme temperature days (Fig. 4). When extreme temperature events occur, CO₂ emissions increase not only due to higher energy consumption for heating or cooling but also because the capacity of renewable energy decreases, further increasing the demand for fossil fuels[1]. High temperatures can reduce the operational efficiency of thermal power plants, and droughts can lower hydropower reservoirs[25-27]. During extreme cold days in winter, the increment of CO₂ emissions ranged from 7.7% (Japan) to 27.1% (Germany) compared to corresponding monthly average CO₂ emissions (Fig. 5). During extreme hot days in summer, the increment of CO₂ emissions ranged from 5.1% (U.S.) to 11.3% (India) compared to corresponding monthly average CO₂ emissions (Fig. 5). Note that the average temperature during summer extreme days in Germany (23.0°C) was below

26°C. Consequently, the increment during summer extreme days was -1.7%. The reconstructed daily $CO_2$ emissions also showed an increase during extreme temperature days. During winter extreme days, the reconstructed dataset showed increments from 1.1% (China) to 20.7% (India), although this result for India is likely affected by the lack of distinction between fossil and biomass-origin CO2 emissions in the underlying data sources. During summer extreme days, except for Germany, the reconstructed dataset showed increments from 0.4% (India) to 3.3% (Japan).

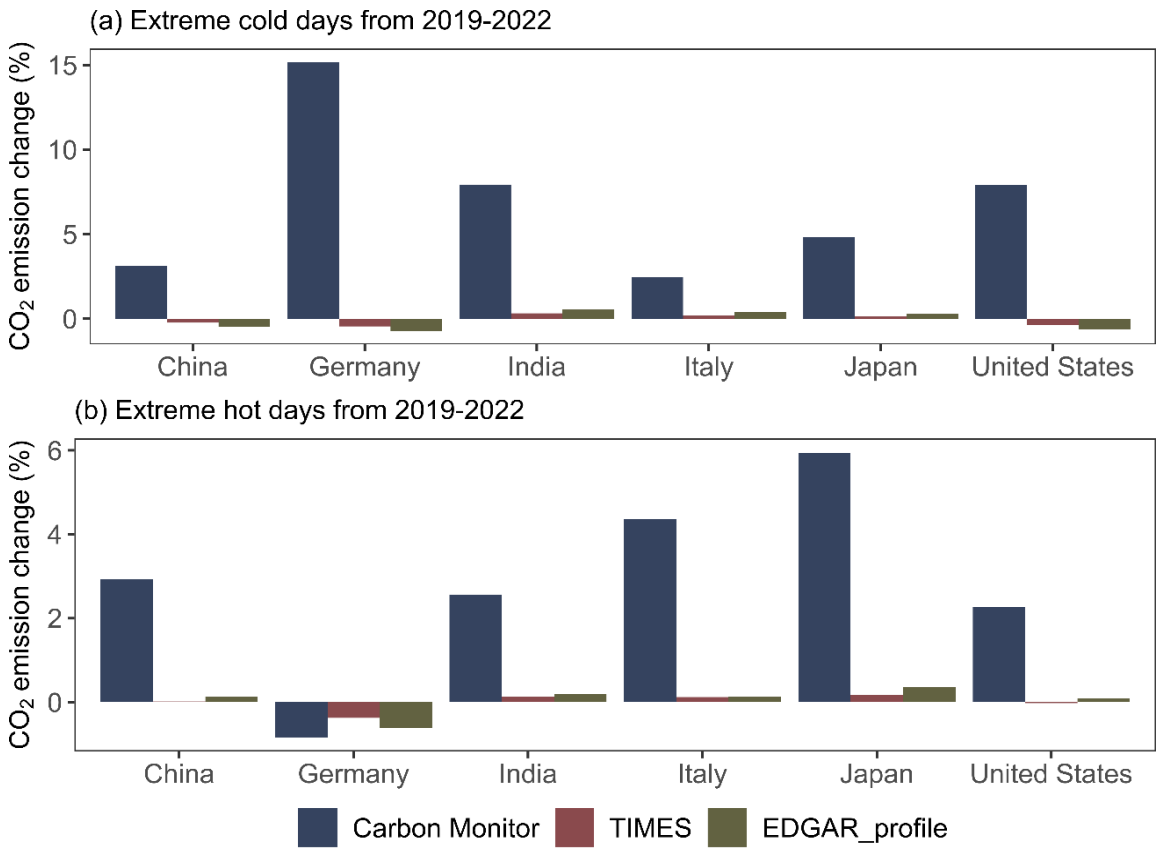

Fig. 5. The average $CO_2$ emission changes (%) (a) during extreme cold days and (b) during extreme hot days from 2019 to 2022 estimated by Carbon Monitor, TIMES, and EDGAR_profile.

After a period of brief decline, the ratio of emissions increment due to temperature extremes to total annual emissions began to increase from the 1990s. In 1970, this ratio was 0.17%, and by 2022, it had reached 0.23% (Fig. 6). The proportion of $CO_2$ emissions increment during extreme hot days consistently increased from 0.02% in 1970 to 0.12% in 2022. Conversely, the increment in $CO_2$ emissions during extreme cold days showed an increasing trend from 1970 to 1980, followed by a decline from 1980 to 2022. The rise between 1970 and 1980 could be associated with global cooling during that period, likely caused by air pollution[28]. Specifically, after 2010, the increase in emissions due to extreme hot events accelerated, corresponding with a rise in the frequency of such events. In contrast, the frequency of extreme cold events gradually decreased. Despite this, the total frequency of extreme temperature events, both hot and cold, significantly increased from 2010 to 2022 compared to 2000-2009. This indicates that global warming has increased the frequency of extreme cold days, while not proportionately reducing the frequency of extreme cold days[29,30].

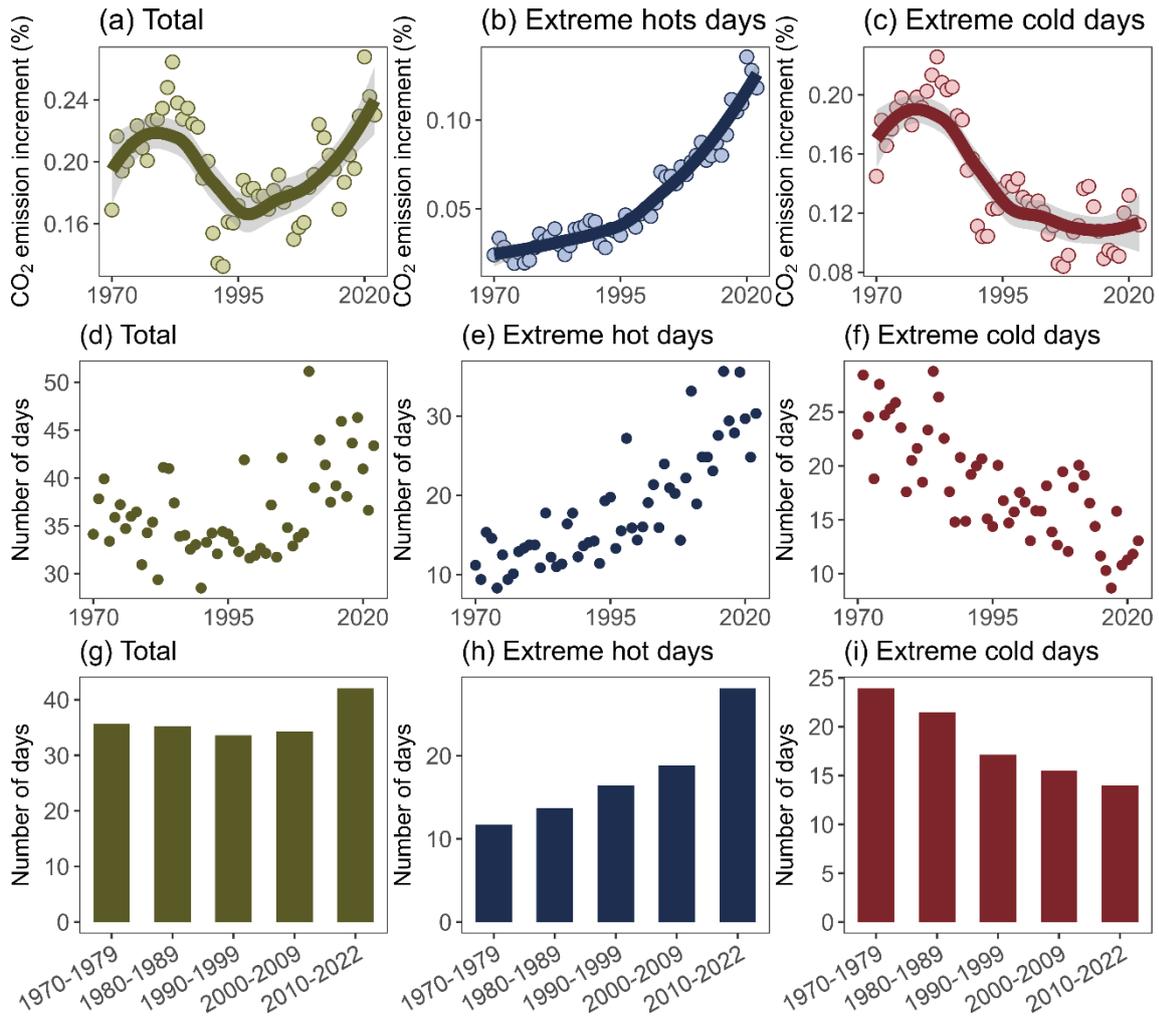

**Fig. 6. Interannual variation in the ratio (%) of emissions increment due to temperature extremes to total annual emissions and frequency (number of days) of temperature extremes.** (a)-(c) represent increment in $CO_2$ emission during extreme cold and hot days, extreme hot days, and extreme cold days. (d)-(f) represent the frequency of extreme cold and hot days, extreme hot days, and extreme cold days. (g)-(i) represent average frequency of extreme cold and hot days, extreme hot days, and extreme cold days in multi-year intervals. The days with population-weighted temperatures falling below the 5th percentile and above the 95th percentile over 53 years (1970-2022) are viewed as extreme cold days and extreme hot days, respectively. Note that increment in $CO_2$ emissions is smoothed by a 5-year moving average. See Methods for the calculation of $CO_2$ emissions changes during extreme cold and hot days.

**Data evaluation**

Following the error propagation equation from the IPCC (See Methods), uncertainty was calculated by combining the uncertainty of EDGAR monthly $CO_2$ emissions and the uncertainty of the reconstruction model. The uncertainty of EDGAR monthly $CO_2$ emissions was evaluated by qualitative analysis[12], while the uncertainty of the reconstruction model was evaluated by quantile regression[31] against the ratio of daily $CO_2$ emissions to the corresponding monthly averages. The uncertainty of global daily $CO_2$ emission was within ±33% based on the rule of error propagation and IPCC 2006 uncertainty analysis[32]. The main source of error arises from the EDGAR monthly $CO_2$ emissions, which are obtained using various proxies such as the heating degree day

(HDD), but also survey data that has been extrapolated beyond its original geographical scope12. While HDD can explain emission changes in the residential sector, it is not applicable when biomass is primarily used for heating. A clear example is India, where fossil $CO_2$ emissions in the residential sector primarily come from cooking with petroleum-based fuels, while most household heating emissions originate from biomass combustion and therefore do not lead to fossil CO2 emissions. Therefore, HDD is unsuitable for indicating emission variations from cooking. Future work should focus on generating improved sub-annual profiles that distinguish between fossil and biogenic $CO_2$ emissions.

We compared our results with other datasets and methods including two emission distribution methods, TIMES[11] and EDGAR_profile[12] and daily $CO_2$ emission data from Vulcan[21] (Table S1). TIMES and EDGAR_profile provide scale factors that reflect daily $CO_2$ emission variations. These scale factors were used to convert EDGAR's monthly $CO_2$ emission estimates into daily values for further comparison with estimates from Carbon Monitor and the reconstructed daily $CO_2$ emissions.

Since the scale factors used to distribute emissions from months to days are fixed, the $CO_2$ emission increments during extreme temperature periods that were estimated by TIMES and EDGAR_profile were not significant (Figs. 5 and S5). TIMES and EDGAR_profile mainly displayed periodic fluctuations related to the day of week. During extreme cold days, TIMES and EDGAR_profile estimated $CO_2$ emission increments ranging from -0.1% (the U.S.) to 0.2% (Italy) and from 0.1% (the U.S.) to 0.3% (Italy), respectively. With regard to extreme hot days, TIMES and EDGAR_profile estimated $CO_2$ emission increments ranging from 0.0% (Italy) to 0.1% (the U.S.) and from -0.1% (Italy) to 0.2% (China), respectively.

Vulcan has used various emission indicators to estimate high temporal resolution $CO_2$ emissions for the U.S. from 2010 to 2015[21]. For the power sector, it obtained real hourly $CO_2$ emissions from the Environmental Protection Agency's Clean Air Markets Division (CAMD) data, which was monitored from emitting stacks. For other sectors, such as residential, where real $CO_2$ emission data was not available, building energy models were applied to simulate $CO_2$ emissions with changing temperatures. The Vulcan emission dataset estimated $CO_2$ emission increments of 4.3% during extreme cold days and 2.8% during extreme hot days, which were close to the estimates from the reconstructed dataset (increments of 5.7% for extreme cold days and 2.2% for extreme hot days; Fig. S6). In contrast, TIMES (0.5% during extreme cold days and 0.0% during extreme hot days) and EDGAR_profile (0.6% during extreme cold days and 0.0% during extreme hot days) displayed relatively lower estimates.

## Discussion

By incorporating the impact of temperature into short-term emission variations, this study improves the estimation of short-term emission changes. Although the impact of temperature on emissions, particularly short-term emission changes, has been widely studied[1,19], the scaling factors provided by TIMES and EDGAR_profile for allocating monthly total $CO_2$ emissions to daily, still do not consider the effect of temperature. The daily $CO_2$ emissions derived by TIMES and EDGAR_profile are widely in atmospheric transport models, meeting the needs for high temporal resolution emission data[12]. Given the ongoing research attempts to use atmospheric transport models to study the impact of extreme temperature on terrestrial carbon sinks[2], the high temporal resolution emission data provided by this study will offer critical data input support.

High temporal resolution $CO_2$ emission datasets support the study of short-term events such as extreme hot events. Based on our reconstructed dataset, we found a trend of increasing emissions due to extreme temperature year by year (Fig. 6). Although similar studies have been conducted[1], this study is the first to examine the long-term effect of temperature on $CO_2$ emissions. In the context of climate change, frequent extreme temperature events significantly impact terrestrial ecosystems[33] and wildfires[34], posing challenges for developing climate mitigation measures. This study reveals the long-term impact of climate change from the perspective of anthropogenic $CO_2$ emissions, showing the emission increase effect, which implies that additional emission reduction measures may be needed to mitigate climate change.

Despite the significant findings, this study has limitations. We only used temperature and time variables for emission reconstruction. Although we achieved good reconstruction performance, and the reconstructed data effectively indicate short-term emission changes, the predictive performance $R^2$ average value based on 10-fold cross-validation reached 0.86, with Russia achieving the highest at 0.97, indicating that the selected temperature and time variables can well simulate daily emission variations. In the future, more environmental factors such as atmospheric $NO_2$, $PM_{2.5}$ concentrations, or more high-resolution data reflecting economic activities, such as stock indices, can be included to improve emission estimates and achieve more accurate high temporal resolution emission estimates.

In addition, the situation in 1970 and now should be different, such as the availability of heating facilities and air conditioning, which leads to a different response of anthropogenic emissions to temperature changes. For example, with widespread air conditioning today, a hot extreme in 2019-2022 must have a different impact on emissions than in the past when air conditioning was not available (similarly, poor insulation in winter and less access to energy by households in the past). Moreover, some hot countries such as India are still a long way from full penetration of air conditioning. In the future, some sectors could be reduced based on socio-economic data such as electrification rate and accessibility to air conditioning to constrain the variability of $CO_2$ emissions in response to temperature changes.

# Method
**Model building**
For reconstructing historical daily $CO_2$ emissions, we first collected the daily $CO_2$ emission estimates from Carbon Monitor[8]. Carbon Monitor, a near-real-time inventory, uses activity data from sectors like power generation and cement production to promptly estimate $CO_2$ emissions, typically shortening the release lag of emission inventories. Given the substantial differences in current $CO_2$ emission levels and the sector distribution compared to historical periods, using recent daily $CO_2$ emissions directly as the response variable in models could lead to concept drift[35]. To mitigate this, we computed the ratio of daily $CO_2$ emissions to their monthly averages as the response variable, as illustrated in Fig. S7. This emission ratio can reflect the daily $CO_2$ emission variations due to the effects of temperature and public holidays. The formula used for this calculation is expressed as follows:

$$Ratio_{c,y,m,i} = \frac{DE_{c,y,m,i,Carbon\ Monitor}}{MAE_{c,y,m,Carbon\ Monitor}}$$

$$MAE_{c,y,m,Carbon\ Monitor} = \frac{\sum_{i=1}^{M} DE_{c,y,m,i,Carbon\ Monitor}}{M}$$

where Daily Emissions, $DE_{y,m,i,Carbon\ Monitor}$, represents the $CO_2$ emissions of country $c$ on day $i$ of month $m$ in year $y$ estimated by Carbon Monitor; Monthly Average Emissions, $MAE_{c,y,m,Carbon\ Monitor}$, is monthly average $CO_2$ emission of country $c$ in month $m$ of year $y$ estimated by Carbon Monitor; $Ratio_{c,y,m,i}$ is emission ratio of country $c$ on day $i$ of month $m$ in year $y$; $M$ is the number of days in month $m$.

Based on the emission ratio, we built machine learning models to simulate the relationship between the emission ratio and predictor variables (i.e., population-weighted temperature and time surrogate variables). We applied eXtreme Gradient Boosting (XGBoost)[20] as the workhorse due to its good interpretability and performance advantages over neural networks and Random Forests when handling tabular data[36-38]. Furthermore, considering the variations in temperature levels and public holidays between countries, we constructed models for each country:

$$\widehat{Ratio_{c,y,m,i}} = f_c(Temp_{c,y,m,i}, X_1, X_2, X_3, X_4)$$

where $f_c$ is the XGBoost model for country $c$; $\widehat{Ratio_{c,y,m,i}}$ is the predicted emission ratio of country $c$ on day $i$ of month $m$ in year $y$; $X_1, X_2, X_3, X_4$ are the predictor variables including the time surrogate variables (i.e., day of week, day of month, month of year, and whether it is a public holiday). The dates of public holidays are obtained using the Python Holidays package. Specifically, in China, the actual Chinese New Year holiday typically extends beyond the officially published period[39]. Therefore, the days from the 23rd day of the 12th lunar month to the 15th day of the first lunar month are also considered public holidays. $Temp_{c,y,m,i}$ represents the population-weighted temperature of country $c$ on day $i$ of month $m$ in year $y$, which is calculated according to following equation:

$$Temp_{c,y,m,i} = \frac{\sum_j Tem_{c,y,m,i,j} \times P_{c,j}}{\sum_j P_{c,j}}$$

where $Temp_{c,y,m,i,j}$ represents the average temperature of $j^{th}$ grid cell in country $c$ on day $i$ of month $m$ in year $y$; $P_{c,j}$ represent the population count in the $j^{th}$ grid cell of country $c$. The population data were sourced from the first edition of the Global Population Count Grid Time Series Estimates[40] and the fourth edition of the Global Grid Population Dataset (GPWv4)[41]. The Global Population Count Grid Time Series Estimates provided population distribution data for 1970-2000 at 10-year intervals, while GPWv4 provided population distribution data for 2000-2020 at 5-year intervals. For the years 1970 to 1999, we used the Global Population Count Grid Time Series Estimates to compute the population-weighted temperature. For the years 2000 to 2022, we used GPWv4. Both datasets have a spatial resolution of 30 arc-seconds, approximately 1 km at the equator. Hourly surface temperatures at 2 meters above ground level were derived from the fifth generation ECMWF reanalysis dataset (ERA5), which has an original spatial resolution of $0.25°$[42]. We employed bilinear interpolation to resample the surface temperature data to a 30 arc-second grid, aligning it with the population data.

We then collected monthly CO₂ emissions estimated by version 8.0 of EDGAR from 1970 to 2018 to derive monthly average CO₂ emissions for each day (monthly CO₂ emissions divided by the number of days in that month). EDGAR, developed by the Joint Research Centre of the European Commission and the Netherlands Environmental Assessment Agency, offers a comprehensive inventory of anthropogenic greenhouse gases and air pollutants[43]. We note that it appears that EDGAR's monthly CO2 emissions estimates represent the sum of fossil and biogenic emissions. Based on the reconstructed emission ratios and monthly average CO₂ emissions estimated by EDGAR for the period 1970-2018, we calculate the daily CO₂ emissions from 1970 to 2018:

$$DE_{c,y,m,i} = MAE_{c,y,m,EDGAR} \times \widehat{Ratio_{c,y,m,i}}$$

$$MAE_{c,y,m,EDGAR} = \frac{MTE_{c,y,m,EDGAR}}{M}$$

where $DE_{y,m,i}$ is the daily CO₂ emissions of country $c$ on day $i$ of month $m$ in year $y$; $MAE_{c,y,m,EDAGR}$ is monthly average CO₂ emissions of country $c$ in month $m$ of year $y$ estimated by EDGAR; $MTE_{c,y,m,EDGAR}$ is monthly total CO₂ emissions of country $c$ in month $m$ of year $y$ estimated by EDGAR. To align with Carbon Monitor, only the CO₂ emissions from those emission categories in EDGAR that correspond to those in Carbon Monitor were used for calculating the monthly averages of CO₂ emissions. Please refer to Table S2 for the correspondence of emission categories between EDGAR and Carbon Monitor[44].

**Model performance evaluation**
We used the 10-fold cross-validation method to validate the model's predictive performance. The CO₂ emission dataset was randomly divided into 10 approximately equal-sized groups. In each iteration, one group was selected for prediction while the remaining nine groups were used for training. This procedure was repeated 10 times, ensuring that each group was used for predictions. We measured the model predictive performance using $R^2$ metric:

$$R^2 = \frac{(\sum_{i=1}^{n} (y_i - \underline{y}) \cdot (\hat{y}_i - \hat{\underline{y}}))^2}{\sum_{i=1}^{n} (y_i - \underline{y})^2 \cdot \sum_{i=1}^{n} (\hat{y}_i - \hat{\underline{y}})^2}$$

where $n$ is the total number of samples; $y_i$ is the observation of sample $i$; $\hat{y}_i$ is the model prediction of sample $i$; $\underline{y}$ is the mean values of observations; $\hat{\underline{y}}$ is the mean values of predictions. Note that we used grid search to optimize the hyperparameters of the XGBoost model[45], selecting those with relatively higher predictive performance and lower computational cost.

**Uncertainty analysis**
The uncertainty linked to each variable contributes to the overall uncertainty in the final result. By applying the error propagation rules and the IPCC 2006 uncertainty analysis method[32], uncertainty was obtained from the following formula:

$$U_{RC,c,y,m,i} = \sqrt{U^2_{EDGAR,c,y,m} + U^2_{Ratio,c,y,m,i}}$$

where $U_{RC,c,y,m,i}$ is the uncertainty for reconstructed daily CO₂ emissions of country $c$ on day $i$ of month $m$ in year $y$; $U_{EDGAR,c,y,m}$ is the uncertainty for EDGAR monthly CO₂ emissions of country $c$ on month $m$ in year $y$; $U_{Ratio,c,y,m,i}$ is the uncertainty for emission ratio of country $c$ on day $i$ of month $m$ in year $y$.

We employed the quantile regression[31] technique to quantify $U_{Ratio,c,y,m,i}$. In quantile regression, the loss function $L(y, \hat{y})$ is expressed as follows:
$$L(y_i, \hat{y}_i) = q \times max^2(0, y_i - \hat{y}_i) + (1 - q) \times max^2(0, \hat{y}_i - y_i)$$
where $q$ indicates the quantile. In this study, $q$ was set to 0.16 and 0.84 to obtain the 16th and 84th percentile forecasts, respectively, forming the [16%, 84%] confidence interval for the predictions. These values were selected so that the prediction interval would ideally cover 68% of the records, corresponding to one standard deviation under the assumption of a Gaussian error distribution[31]. Subsequently, the $U_{Ratio}$ was obtained according to following equation:
$$U_{Ratio} = \frac{(\widehat{y_{i,upper}} - \widehat{y_{i,lower}})}{2 \times \widehat{y_{i,middle}}}$$
where $U_{Ratio}$ is uncertainty for emission ratio; $\widehat{y_{i,lower}}$, $\widehat{y_{i,middle}}$, and $\widehat{y_{i,upper}}$ are values predicted for $y_i$ where $q$ is set to 0.16, 0.5, and 0.84, respectively.

**Calculate the changes in CO₂ emissions during extreme temperature days**
Since $CO_2$ emissions exhibit significant seasonal variation, we need to choose an appropriate emission baseline to calculate relative changes as the metrics for quantifying the impact of extreme events. We quantified the impact of extreme events on $CO_2$ emissions according to the following equations:
$$Rec_{c,y,m,i} = DE_{c,y,m,i} - MAE_{c,y,m}$$
where $Rec_{c,y,m,i}$ is the relative changes in $CO_2$ emissions of country $c$ on day $i$ of month $m$ in year $y$; $DE_{c,y,m,i}$ is the daily of country $c$ on day $i$ of month $m$ in year $y$; $MAE_{c,y,m}$ is the monthly average $CO_2$ emissions of country $c$ on month $m$ in year $y$.

# Supplementary Material
**Reconstructing Global Daily CO$_2$ Emissions via Machine Learning**

**Table S1. The fundamental information on the dataset or method used by this study.**

| Dataset/Method | Carbon Monitor | EDGAR | Vulcan | TIMES | EDGAR_profile |
|---|---|---|---|---|---|
| Spatial resolution | Country | Country | 1km×1km | 0.25°×0.25° | Country |
| Temporal resolution | Daily | Monthly | Hourly | Daily | Daily |
| Temporal coverage | 2019-2022 | 1970-2018 | 2010-2015 | - | - |
| Version | - | V8 | V3.0 | - | r1 |
| Reference | 1 | 2 | 3 | 4 | 5 |

Note that each dataset or method may provide multiple products with different resolutions, but only the information on the products used in this study is displayed. TIMES and EDGAR_profile are emission allocation methods, and their temporal resolution of daily means that they can distribute $CO_2$ emissions from monthly to daily.

**Table S2.** The correspondence between EDGAR's emission categories, specified by the IPCC 1996 code, and Carbon Monitor's emission categories.

| IPCC 1996 code | Description | Carbon Monitor |
|---|---|---|
| 1A1a | Public electricity and heat production | Power |
| 1A1bc | Other Energy Industries | Industry (incl. Cement Process) |
| 1A2 | Manufacturing Industries and Construction | Industry (incl. Cement Process) |
| 2A1 | Cement production | Industry (incl. Cement Process) |
| 1A3a | Domestic aviation | Domestic aviation |
| 1A3b | Road transportation no resuspension | Ground Transport |
| 1A3c | Rail transportation | Ground Transport |
| 1A3d | Inland navigation | Ground Transport |
| 1A3e | Other transportation | Ground Transport |
| 1A4 | Residential and other sectors | Residential |
| 1A5 | Other Energy Industries | Residential |
| 1C2 | Memo: International navigation | International shipping |
| 1C1 | Memo: International aviation | International aviation |

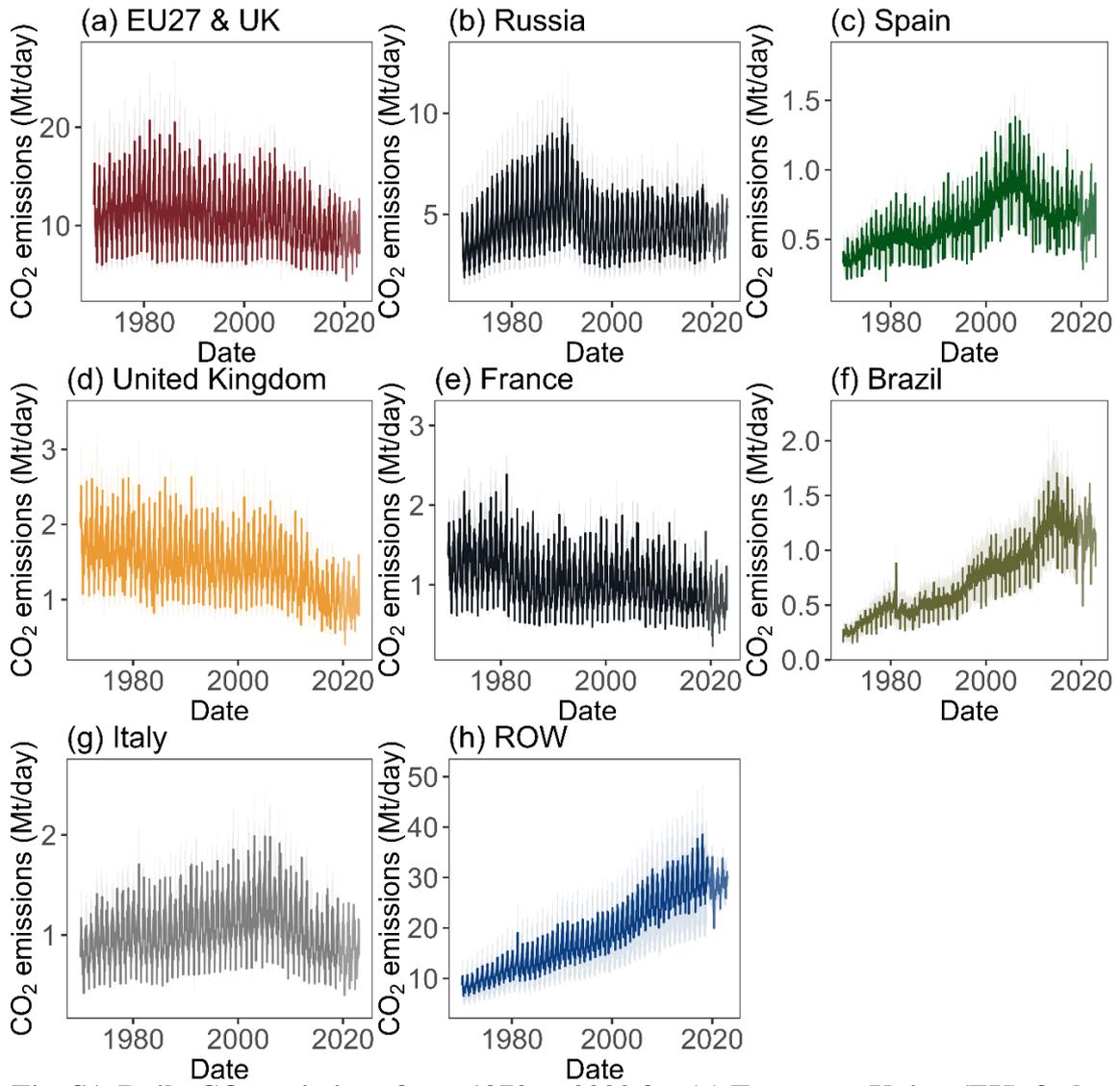

**Fig. S1. Daily CO₂ emissions from 1970 to 2022 for (a) European Union (EU) & the United Kingdom (UK), (b) Russia, (c) Spain, (d) the United Kingdom, (e) France, (f) Brazil, (g) Italy, and (h) rest of world (ROW).** The shaded areas indicate the uncertainty of $CO_2$ emissions.

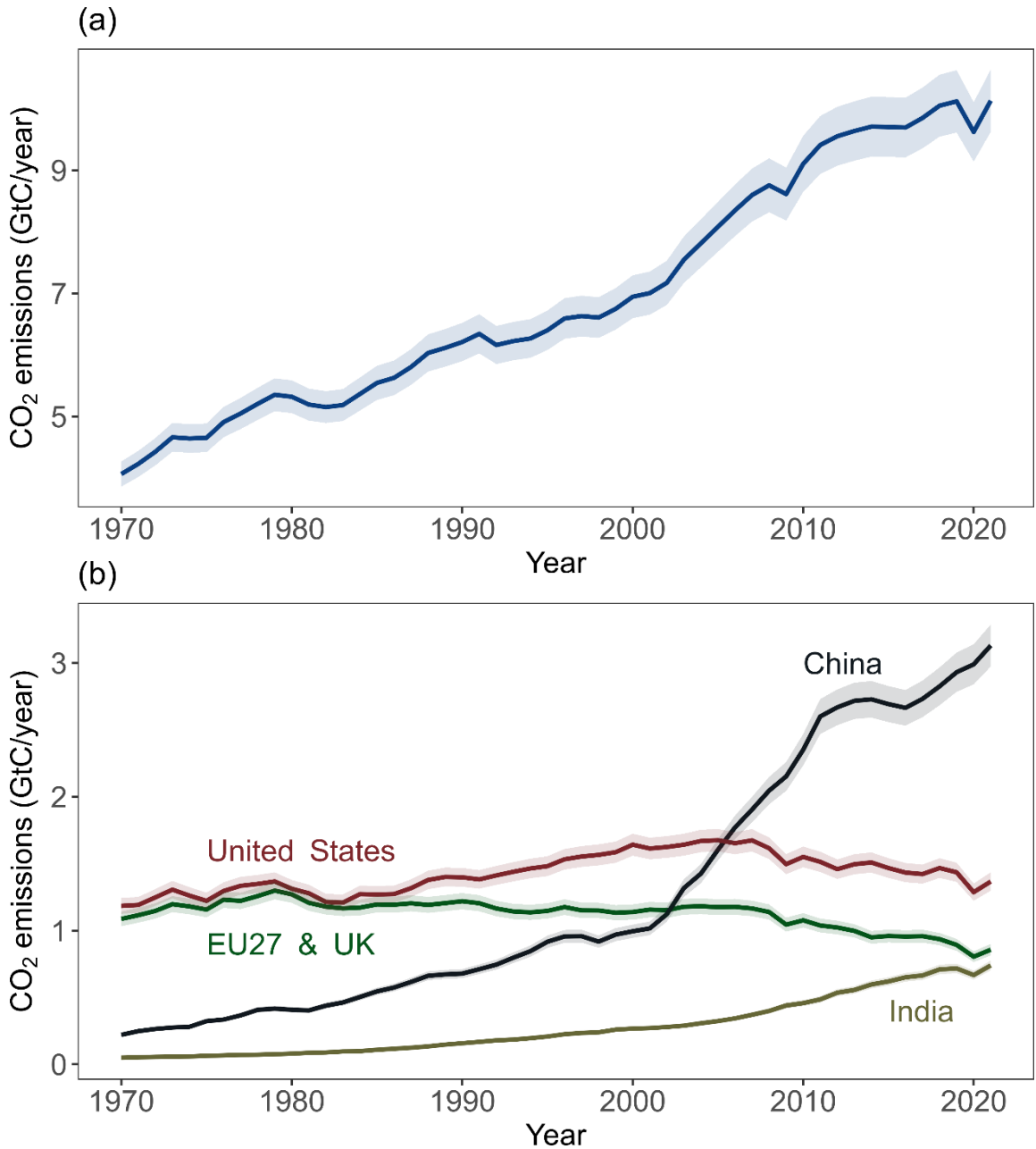

**Fig. S2. Annual CO₂ emissions from 1970 to 2021 for (a) global and (b) individual countries estimated by Global Carbon Budget 2022[6].** The shaded areas indicate the uncertainty of $CO_2$ emissions.

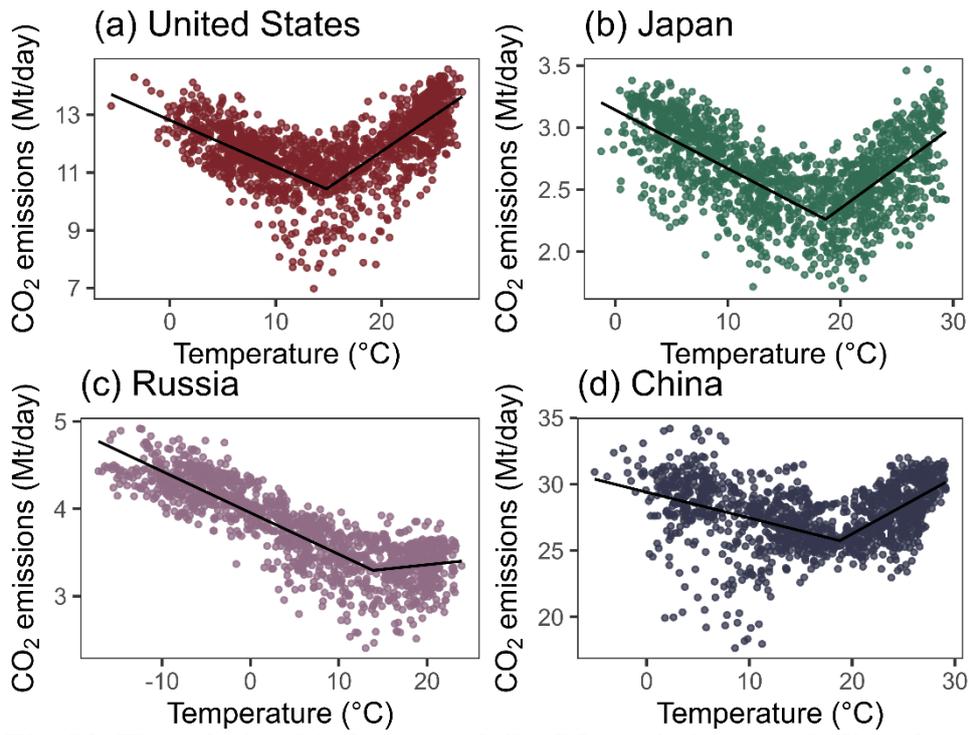

**Fig. S3. The relationship between daily CO$_2$ emissions (excluding the residential sector) and population-weighted temperature for (a) the United States, (b) Japan, (c) Russia, and (d) China.** Note that in (g-l), data records for public holidays have been removed.

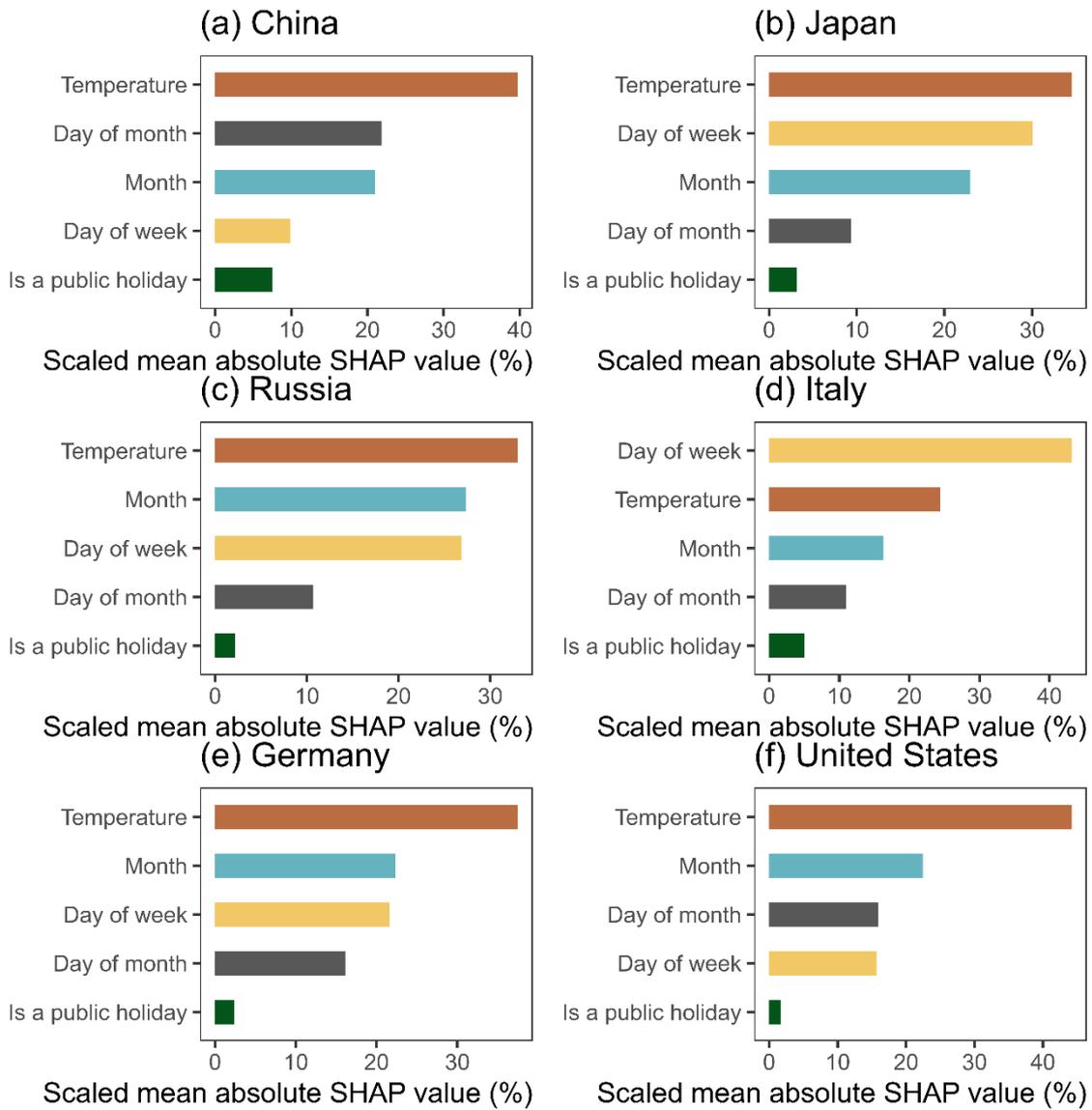

**Fig. S4. Variable contribution to daily $CO_2$ emissions measured by scaled mean absolute SHAP value.**

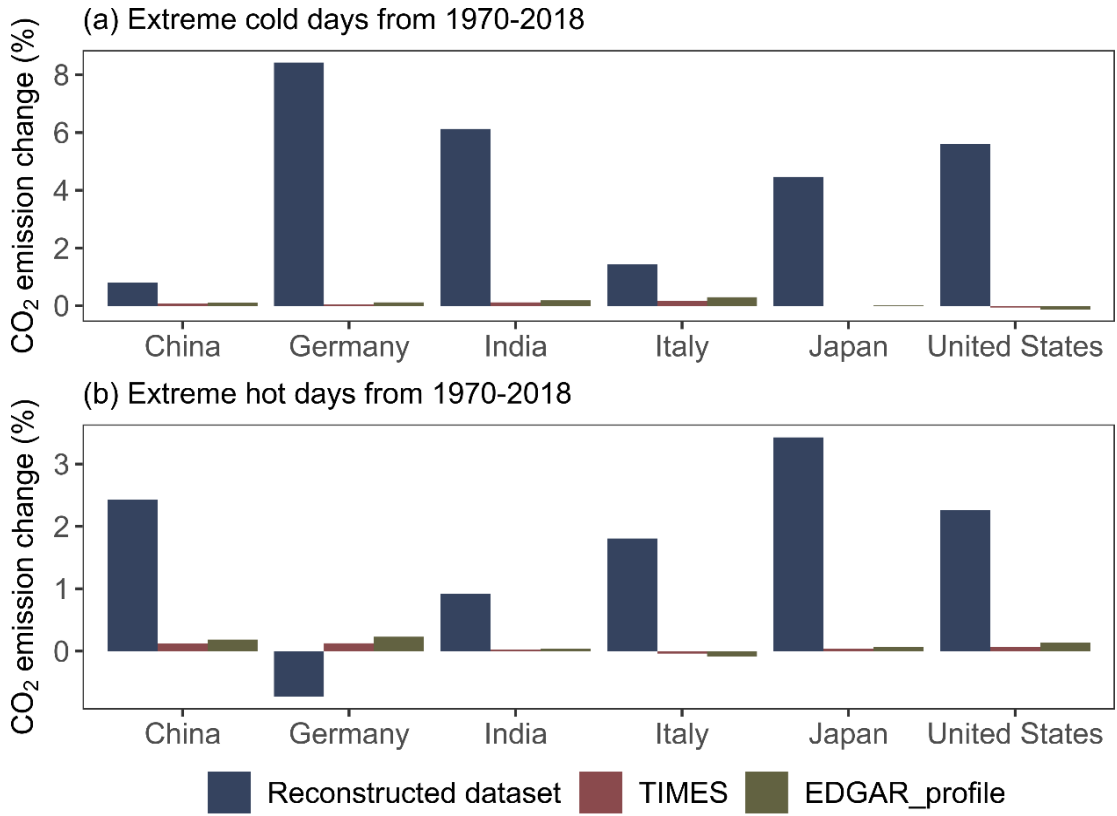

Fig. S5. The average $CO_2$ emission increments (%) (a) during extreme cold days and (b) during extreme hot days from 1970 to 2018 estimated by the reconstructed dataset, TIMES, and EDGAR_profile.

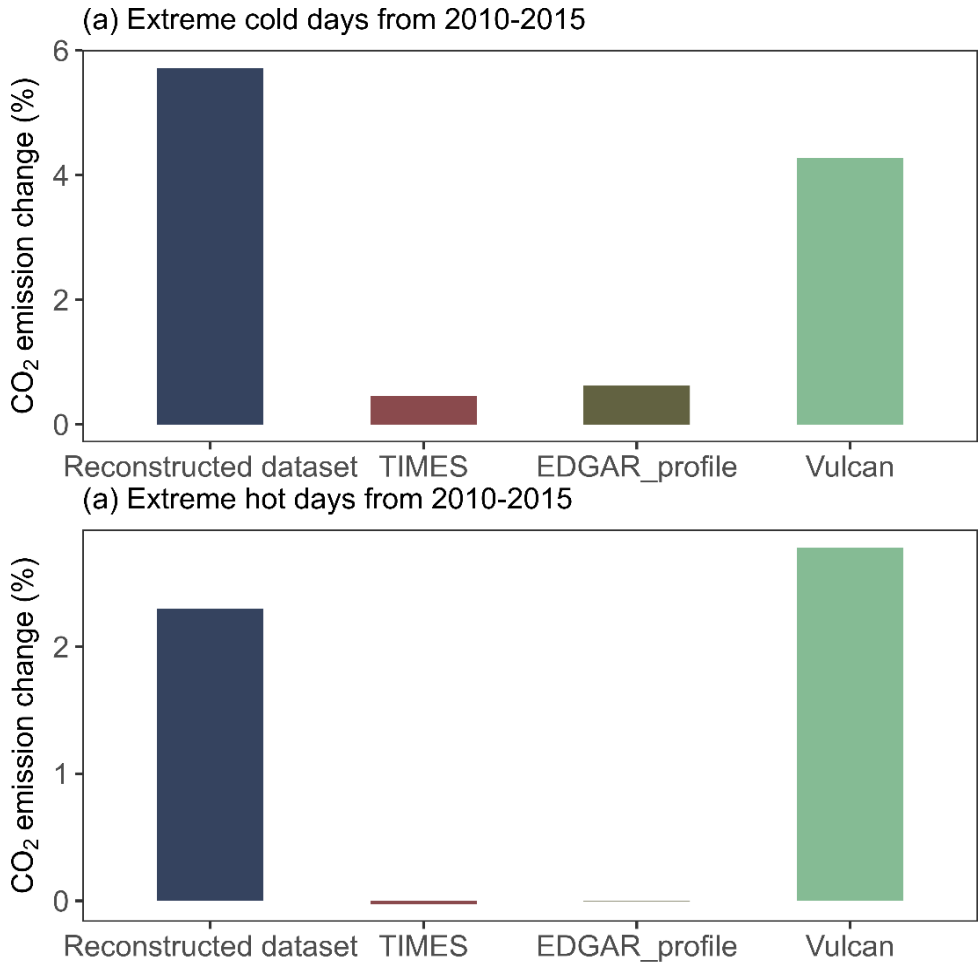

Fig. S6. The average $CO_2$ emission increments (%) (a) during extreme cold days and (b) during extreme hot days from 2010 to 2015 estimated by the reconstructed dataset, TIMES, EDGAR_profile, and Vulcan in the United States.

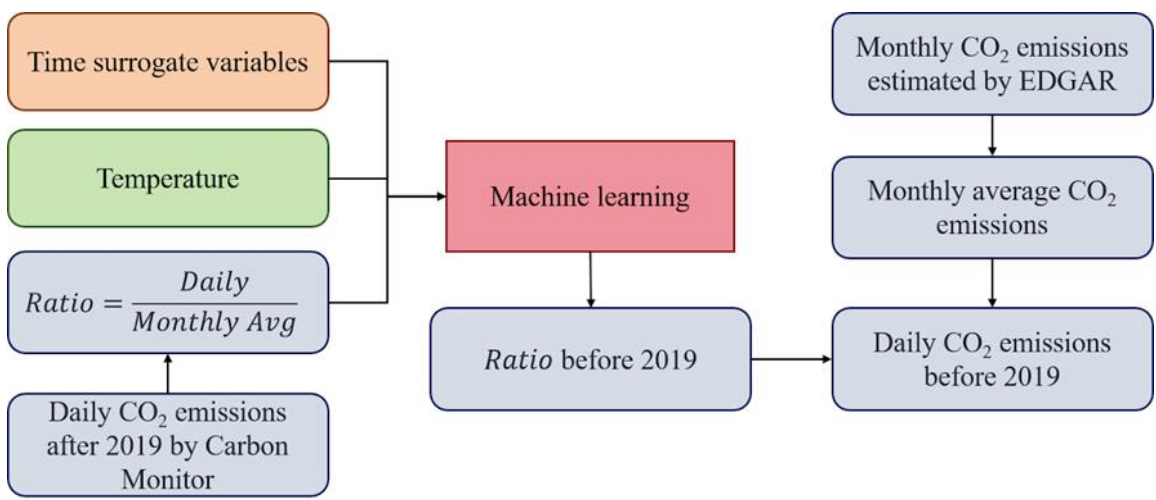

Fig. S7. Flowchart of the algorithm used for reconstructing historical daily $CO_2$ emissions based on machine learning.